\documentstyle[aps,preprint,tighten]{revtex}

\draft

\begin{document}

\title{When the Casimir energy is not a sum of zero-point energies}
\author{Luiz C.\ de Albuquerque\footnote{E-mail:
lclaudio@fatecsp.br}}
\address{Faculdade de Tecnologia de S\~ao Paulo - CEETEPS - UNESP \\
Pra\c{c}a Fernando Prestes, 30, 01124-060 S\~ao Paulo, SP, Brazil} 
\author{R.\ M.\ Cavalcanti\footnote{E-mail: rmoritz@if.ufrj.br}}
\address{Instituto de F\'{\i}sica, Universidade Federal 
do Rio de Janeiro \\
Caixa Postal 68528, 21945-970 Rio de Janeiro, RJ, Brazil}
\date{3 September 2001}
\maketitle

\begin{abstract}

We compute the leading radiative correction to the 
Casimir force between two parallel plates in the 
$\lambda\Phi^4$ theory. Dirichlet and periodic boundary 
conditions are considered. A heuristic approach, in 
which the Casimir energy is computed as the sum of one-loop 
corrected zero-point energies, is shown to yield 
incorrect results, but we show how to amend it. The 
technique is then used in the case of periodic boundary 
conditions to construct a perturbative expansion which 
is free of infrared singularities in the massless 
limit. In this case we also compute the next-to-leading 
order radiative correction, which turns out to be proportional to 
$\lambda^{3/2}$.

\end{abstract}




\section{Introduction}

An important question in quantum field theory is the response of
the vacuum fluctuations to perturbations of the space-time
manifold: in the absence of a consistent quantum theory of
gravitation in four space-time dimensions one is led to study vacuum
fluctuations of matter or gauge fields in the presence of an
external (i.e., classical) gravitational field \cite{Fulling}.
One may also ask how the properties of a field theory are 
affected by the topology of space-time or by the
presence of boundaries, which impose constraints on the fields.
For instance, periodic boundary
conditions on a spatial sector are a key ingredient in
compactification schemes of Kaluza-Klein theories \cite{KK}.
Boundary conditions (BC) are also used to describe complicated
physical systems in a simplified mathematical framework. In the
electromagnetic Casimir effect \cite{Casimir} one considers
classical conductor plates (perfect mirrors), with the field
satisfying Dirichlet BC on them. The
analogous condition in the MIT bag model is the perfect
confinement of quarks and gluons to the interior of hadrons
\cite{MIT}. In thermal field theory, periodic or anti-periodic BC
in the imaginary-time are the starting point of the Matsubara
formalism \cite{Thermal}. Finally, the study of surface effects on
the critical properties of a (magnetic, binary liquid, etc.) system
leads in many cases to the analysis of scalar field
theories subject to certain boundary conditions \cite{Diehl}.

Although BC have been extensively studied in quantum field 
theory models, there remains a lot of questions to be answered. 
In this paper we will investigate some unusual features of 
periodic and Dirichlet BC on one spatial coordinate. In the 
remainder of the Introduction we will give some motivations 
to the study of these particular types of boundary conditions. 

Quantum field theories in compactified spaces (i.e., with
periodic boundary conditions in some spatial directions)
have been the subject of considerable interest in the literature
\cite{Ford,Toms,Periodic}. The calculation of the effective potential
in spontaneously broken symmetry theories shows that the
compactification process may introduce a mechanism for dynamical
symmetry restoration. Generation of a dynamical mass is connected
with the inclusion of a new scale, the compactification
radius $R$.

There is a complete mathematical analogy between
compactified field theory and thermal field theory (TFT)
in the Matsubara formalism. In the latter, the inverse
temperature $\beta=1/T$ is the compactification radius
in the imaginary-time direction.  The well-known fact that
thermal effects do not lead to new ultraviolet divergences in TFT
(besides the usual ones found at $T=0$) \cite{Thermal} applies as well
to compactified field theories. On the other hand,
the infrared properties of the TFT are very different from the ones
at zero temperature. The free energy of massless $\lambda\Phi^4$
theory in three spatial
dimensions develops new infrared divergences at order $\lambda^2$
in perturbation theory\cite{Thermal}. The dominant infrared
divergences come from the $n=0$ mode of the loop momenta.
A proper treatment of the collective effects leads to a
correction of order $\lambda^{3/2}$ to the free energy.

The infrared behavior of the compactified field theory mimics the
one at finite temperature, at least in the case of one spatial
compactified coordinate. In a perturbative treatment, the $n=0$
mode generates new infrared divergences in the compactified
version of the $\lambda\Phi^4$ theory.
This seems to be not so
well-appreciated in the literature. To fill this gap, we apply the
resummation method developed by Braaten, Pisarski and others 
(in the context of TFT) \cite{Braaten,RRParwani,Frenkel} 
to the compactified $\lambda\Phi^4$ theory.

Symmetries in quantum field theory put very stringent conditions
on the perturbative renormalization of a model and in its
physical predictions. Lorentz invariance (rotational invariance
in the Euclidean case) is of paramount
importance in this respect. However, external conditions or
dynamical effects may lead to its breakdown. There
is a growing interest on effective field theories in which
this occurs (e.g., non-commutative field theories
\cite{No-Comut}, anisotropic systems \cite{Lifshitz}, and Chern-Simons
theories \cite{CS}). 
Theories defined in finite volumes or in the presence of
macroscopic bodies (as in the Casimir effect)
may provide useful insights on the consequences of lack of
Lorentz symmetry to the renormalization program.

Recently, there has been much effort in the computation 
of radiative corrections to the Casimir energy,
specially in QED \cite{RCCF}. 
One of the purposes of this paper is to discuss an alternative
method to compute such corrections. 
For simplicity we work with
the $\lambda\Phi^4$ theory subject to
Dirichlet boundary conditions on a pair of parallel
plates. The method is based on a resummation of the 
perturbative series for the two-point Green function,
and leads to a Klein-Gordon equation
in which the one-loop self-energy acts as an effective
scalar potential. In four space-time 
dimensions this equation can be solved exactly 
in the massless case.
The new set of (resummed) eigenvalues contain
radiative corrections of all orders in $\lambda$,
and reduce to the free ones for $\lambda=0$.
The computation of the sum of effective zero-point energies,
including non-perturbative corrections and renormalization
issues, is discussed in detail.

The plan of the paper is as follows. In Section \ref{DBC} 
we fix the
conventions and discuss the resummation technique
in the $\lambda\Phi^4$ theory with Dirichlet boundary
conditions. We solve the effective Klein-Gordon equation,
and obtain the ``improved'' eigenvalues. The solution is used to
obtain the resummed Casimir energy, including radiative
corrections. In Section \ref{PBC} we discuss the resummation
of the vacuum energy in the case of periodic boundary conditions;
this sheds some light on the results of Section \ref{DBC}.
In the Conclusion we discuss the drawbacks of this method
as well as other minor points. Finally, three Appendices collect
some mathematical results used in the paper.


\section{Dirichlet boundary conditions}
\label{DBC}

Boundary conditions breaking  the full Lorentz invariance in
general pose new problems to the renormalization program. For
some geometries and boundary conditions (depending also on the
spin of the field) it may be
necessary to introduce surface counterterms besides the
bulk ones. For instance, in the MIT Bag
model the free energy is ill-defined at one-loop due to an extra
singularity which shows up as the surface is approached
\cite{Milton}. The
standard recipe associates a free parameter to each distinct
singular term, included as a new counterterm in the starting
Lagrangian. If this procedure continues to all orders, with the
consequent loss of predictive power, we say that the theory is
non-renormalizable due to the boundary conditions.

In a remarkable paper, Symanzik gave strong arguments showing the
renormalizability of the $\Phi^4$ theory in the presence
of flat boundaries \cite{Sy}. In particular, he showed that the
renormalized Casimir pressure for disjoint boundaries and Dirichlet
BC is finite to all orders in perturbation theory.
He also verified explicitly that no surface counterterms
are needed in the computation of the two-loop
vacuum energy. $\Phi^4$-type theories are still renormalizable 
for more general boundary conditions and surfaces,
but at the price of introducing surface counterterms \cite{Diehl,Sy}.
We wish to point out
here that many proposals have been made in order to avoid the
surface-like singularities. These include, among others,
the ``softening''
of the Dirichlet BC \cite{Act,LC} or treating the boundaries
as quantum mechanical objects with a nonzero position
uncertainty \cite{FN}. However, this question is
outside the scope of our present discussion.

A key ingredient in our computation of the Casimir
energy is the self-energy of the field, as it determines the
shift in the single-particle energy levels. Since there
is some disagreement among existent results in the 
literature \cite{Ford,Toms,Sy,Nami}, we present its
computation in some detail. 


\subsection{Self-energy}

We work in $D=(d+1)+1$ dimensional Minkowski space-time, and define
$x^\mu\equiv (t,{\bf x},z)$, with ${\bf x}=(x^1,\ldots,x^d)$.
The renormalized Lagrangian reads
[$\hbar=c=1$, $\eta_{\mu\nu}={\rm diag}(+,-,...,-)$]
\begin{equation}\label{1}
{\cal L} = {\cal L}_0 + {\cal L}_I
=\left\{\frac{1}{2}\,(\partial\Phi)^2-\frac{1}{2}\,m^2\Phi^2\right\}
+\left\{-\frac{\lambda}{4!}\,\Phi^4+{\cal L}_{\rm ct}\right\},
\end{equation}
with ${\cal L}_{\rm ct}$ the counterterm Lagrangian.

We impose Dirichlet BC on a pair of plates at
$z=0$ and $z=\ell$: $\Phi(z=0)=\Phi(z=\ell)=0$.
The bare Feynman propagator with
Dirichlet BC may be written as an expansion in multiple
reflections \cite{Balian}:
\begin{equation}\label{29}
\Delta_F(x,x')=\sum_{n=-\infty}^{\infty}\left[
\Delta_F^{(0)}(x_n-x_+')-\Delta_F^{(0)}(x_n-x_-')\right],
\end{equation}
where $x_n=(t,{\bf x},z+2n\ell)$,
$x_{\pm}'=(t',{\bf x}',\pm z')$, and
$\Delta_F^{(0)}$ is the bulk free propagator, which
for $D>2$ is given by \cite{BS}
\begin{equation}\label{30}
\Delta_{F}^{(0)}(x)=\frac{1}{(2\pi)^{D/2}}
\left(\frac{m}{\sqrt{-x^2+i\epsilon}}\right)^{(D-2)/2}
K_{(D-2)/2}\left(m\sqrt{-x^2+i\epsilon}\right),
\end{equation}
with $\sqrt{-x^2+i\epsilon}=i\sqrt{x^2}$ if $x^2>0$
($\epsilon\to 0^+$).
Actually, what we are interested in is
$\Delta_F(x,x)$. It follows from Eq.\ (\ref{29}) that
\begin{equation}\label{31}
\Delta_{F}(x,x)=\sum_{n=-\infty}^{\infty}\left[
\Delta_F^{(0)}(2n\ell)-\Delta_F^{(0)}(2z+2n\ell)\right].
\end{equation}
The term $\Delta_{F}^{(0)}(0)$ contains the usual UV
singularity. It can be removed, as usual, by a mass
renormalization.

In the massless case the bulk free propagator
gets simplified, and it is possible to
find a closed expression for
$\Delta(x)\equiv\Delta_F(x,x)-\Delta_F^{(0)}(0)$. Using
\begin{equation}
\Delta_{F}^{(0)}(x;m=0)=\frac{\Gamma\left(\frac{D}{2}-1\right)}
{4\pi^{D/2}}\,|x|^{2-D},
\end{equation}
one finds, for $D=4$,
\begin{equation}\label{pre32}
\Delta(x;m=0)=\frac{1}{16\pi^2 \ell^2}\left[
2\,\psi'(1)-\psi'\left(\frac{z}{\ell}\right)
-\psi'\left(1-\frac{z}{\ell}\right)\right],
\end{equation}
where $\psi(x)$ is the digamma
function \cite{Grad}. Eq.\ (\ref{pre32}) can be
simplified to
\begin{equation}\label{32}
\Delta(x;m=0)=
-\frac{1}{16\ell^2}\left[\csc^2\left(\frac{\pi z}{\ell}\right)
-\frac{1}{3}\right]\qquad(D=4).
\end{equation}

Let us take a closer look at the $D=3$ case, keeping $m\neq 0$. We
obtain from Eqs.\ (\ref{30}) and (\ref{31}), after
changing the summation variables and using the explicit form of
$K_{1/2}(x)$,
\begin{equation}\label{33}
\Delta(x)=\frac{1}{8\pi\ell}\left[
2e^{-2m\ell}S(2m\ell,1)
-e^{-2mz}S\left(2m\ell,\frac{z}{\ell}\right)
-e^{-2m(\ell-z)}S\left(2m\ell,1-\frac{z}{\ell}\right)\right],
\end{equation}
where
\begin{equation}\label{36}
S(a,b)\equiv \sum_{n=0}^\infty \frac{e^{-an}}{n+b}.
\end{equation}
The massless limit must be taken with care, as each of
the series in Eq.\ (\ref{33}) is logarithmically divergent.
As we shall show, the divergent terms cancel
in the complete formula (\ref{33}).
Indeed, the asymptotic limit of $S(2m\ell,b)$ as $m\to 0$
is given by
\begin{eqnarray}\label{36b}
S(2m\ell,b)&=&\sum_{n=0}^\infty e^{-2m\ell n}\left(\frac{1}{n+b}
-\frac{1}{n+1}\right)
+\sum_{n=0}^\infty \frac{e^{-2m\ell n}}{n+1}
\nonumber \\
&\stackrel{m\to 0}{\sim}&-\gamma-\psi(b)-\ln(2m\ell)+O(m\ell),
\end{eqnarray}
where $\gamma=0.577\ldots$ is the Euler constant.
The logarithmic terms cancel in Eq.\ (\ref{33}), so
that we can now take the limit $m\to 0$ safely, obtaining
\begin{equation}\label{39}
\Delta(x;m=0)= \frac{1}{8\pi \ell}\left[2\gamma
+\psi\left(\frac{z}{\ell}\right)
+\psi\left(1-\frac{z}{\ell}\right)\right]\qquad(D=3).
\end{equation}

The renormalized one-loop self-energy is given by
$\Sigma^{(1)}(x)=\frac{\lambda}{2}\Delta_F(x,x)+
\delta m^2$. The mass counterterm is fixed by the condition
$\lim_{\ell\to \infty}\Sigma^{(1)}=0$. This amounts to
remove the contribution of the bulk free propagator
from Eq.\ (\ref{31}). With this choice of mass renormalization
we have $\Sigma^{(1)}(x)=\frac{\lambda}{2}\Delta(x)$.
Therefore, the self-energy is infrared finite in the
massless case also at $D=3$, in disagreement with 
Ref.\ \cite{Nami}.
(However, it is infrared divergent in the case of
Neumann boundary conditions. 
The propagator is then given by Eq.\ (\ref{29})
with the minus sign on its right-hand side (RHS) 
replaced by a plus sign.
As a consequence, the logarithmic terms in the massless
limit of $\Delta(x)$ do not cancel.)

From now on, we shall focus our attention on the
massless case at $D=4$.


\subsection{Radiative corrections to the Casimir energy:
a heuristic approach}

Computations of the Casimir energy in the literature are
restricted to perturbation theory. A non-perturbative calculation
would be a very interesting result. Our goal here is more modest.
We will discuss the computation of the Casimir energy in the
approximation where the two-point Green function is dressed with
an arbitrary number of insertions of the one-loop self-energy
(``daisy'' resummation). This approximation contains the leading
correction in a $1/N$ expansion. (In our case, however,
$N=1$. With this caveat in mind, let us proceed.) The
Casimir energy will be given formally by
\begin{equation}\label{Eformal}
E=\frac{1}{2}\sum_{\alpha}\omega_{\alpha},
\end{equation}
where $\omega_{\alpha}$ are the positive poles of the dressed
two-point function $\widetilde{G}^{(2)}$ in the
frequency domain.

To compute $\widetilde{G}^{(2)}$ we note that it satisfies
\begin{equation}\label{42}
\left[\partial_x^2+\Sigma^{(1)}(x)\right]
\widetilde{G}^{(2)}(x,x')=-i\,\delta^{(4)}(x,x').
\end{equation}
As usual, the solution to Eq.\ (\ref{42}) can be written as
\begin{equation}\label{tG2}
\widetilde{G}^{(2)}(x,x')=-i\sum_{\alpha}\frac{
\phi_{\alpha}^{}(x)\,\phi_{\alpha}^*(x')}
{\Lambda_{\alpha}},
\end{equation}
where $\Lambda_{\alpha}$ and $\phi_{\alpha}(x)$ are the
eigenvalues and (normalized) eigenfunctions,
respectively,
of the Klein-Gordon operator $\partial^2+\Sigma^{(1)}$:
\begin{equation}\label{43}
\left[\partial^2+\Sigma^{(1)}(x)\right]
\phi_{\alpha}(x)=\Lambda_{\alpha}\,\phi_{\alpha}(x).
\end{equation}
Since $\Sigma^{(1)}(x)$ is a function of $z$ alone,
we can reduce the above equation to an ordinary
differential equation by writing
$\phi(x)=e^{-i\omega t+i{\bf p}\cdot{\bf x}}\,\varphi(z)$:
\begin{equation}\label{44}
\left\{-\frac{d^2}{dz^2}-\sigma^2-\frac{\pi^2g}{\ell^2}
\left[\csc^2\left(\frac{\pi z}{\ell}\right)-\frac{1}{3}
\right]\right\}\varphi(z)=0,
\end{equation}
where $\sigma^2\equiv\Lambda+\omega^2-{\bf p}^2$ and
$g\equiv\lambda/32\pi^2$. Now we make the change of
variable $z=\ell y/\pi$ and get
\begin{equation}\label{EqSch}
\left(\frac{d^2}{dy^2}+k^2+\frac{g}{\sin^2y}\right)
\varphi(y)=0\qquad\left(k^2\equiv\frac{\ell^2\sigma^2}{\pi^2}
-\frac{g}{3}\right).
\end{equation}

Equation (\ref{EqSch}) may be viewed as the time-independent
Schr\"odinger equation for a particle of mass
$\widetilde{m}=1/2$
moving in the potential $V(y)=-g\,\csc^2y$
(inverted Poschl-Teller), with energy $E=k^2$.
Its solution is discussed in Appendix \ref{AppendixA}. In
particular, it is shown that $k^2=(n+s)^2$ ($n=1,2,...$),
with $s\equiv\frac{1}{2}\left(-1+\sqrt{1-4g}\right)$. From the
definition of $k^2$ and $\sigma^2$ it follows that the
eigenvalues of the Klein-Gordon operator have the form
\begin{equation}\label{Lambda}
\Lambda=-\omega^2+{\bf p}^2+\frac{\pi^2}{\ell^2}
\left[(n+s)^2+\frac{g}{3}\right]\qquad(n=1,2,\ldots).
\end{equation}
{}From Eqs.\ (\ref{tG2}) and (\ref{Lambda}) it follows
that the (positive) poles of $\widetilde{G}^{(2)}$ are
given by
\begin{equation}\label{omega_n}
\omega_n({\bf p})
=\sqrt{{\bf p}^2+\frac{\pi^2}{\ell^2}\left[
(n+s)^2+\frac{g}{3}\right]}\qquad(n=1,2,\ldots).
\end{equation}

Before we proceed with the calculation of the Casimir energy,
a remark is in order here. As we have seen,
the (renormalized) one-loop self-energy is a function of $x$.
It may be tempting to interpret $\Sigma^{(1)}(x)$ (more
generally, $m^2+\Sigma^{(1)}(x)$)
as a position-dependent (squared) mass of the field.
A problem would then occur in regions where
$\Sigma^{(1)}(x)<0$, for this could imply the
presence of tachyons in the theory.
For that reason, Ford and Yoshimura \cite{Ford} argued
that models which exhibits this behavior (such as the one
we are considering) are unphysical. However, the analysis of equation
(\ref{44}), summarized in Appendix \ref{AppendixA},
shows that its solutions do not have imaginary frequencies
as long as $\lambda<\lambda_{\rm crit}= 8\pi^2$
(which is anyway well outside the range of validity
of perturbation theory). The one-loop effective theory
is consistent in this case. On the other hand,
for $\lambda>8\pi^2$ the Schr\"odinger equation (\ref{44}) leads to
an energy spectrum which is unbounded from below,
rendering the associated effective field theory ill-defined.
This solves a long-standing problem of interpretation.

Substituting the eigenfrequencies (\ref{omega_n}) into
Eq.\ (\ref{Eformal}) we obtain
the following expression for the
Casimir energy per unit area:
\begin{equation}\label{48}
{\cal E}
=\frac{1}{2}\,\mu^{1-2\nu}\sum_{n=1}^\infty\int
\frac{d^2p}{(2\pi)^2}\left\{{\bf p}^2+\frac{\pi^2}{\ell^2}
\left[(n+s)^2+\frac{g}{3}\right]\right\}^{\nu}
\Bigg|_{\nu=1/2}.
\end{equation}
The formal sum over zero-point energies has been
analytically regularized; we shall set $\nu=1/2$
at the end of the calculation. The factor $\mu^{1-2\nu}$,
where $\mu$ is a mass parameter,
keeps the RHS of Eq.\ (\ref{48})
with the dimension of energy per unit area.

Integrating Eq.\ (\ref{48}) over ${\bf p}$, we get
\begin{equation}\label{49}
{\cal E}
=\frac{\mu^{1-2\nu}}{8\pi}\,
\frac{\Gamma(-\nu-1)}{\Gamma(-\nu)}
\left(\frac{\pi}{\ell}\right)^{2(\nu+1)}
{\cal H}\left(-\nu-1;s,\frac{g}{3}\right),
\end{equation}
where the function ${\cal H}(z;s,a^2)$ is defined as
\begin{equation}\label{calH}
{\cal H}(z;s,a^2)\equiv\sum_{n=1}^{\infty}
\left[(n+s)^2+a^2\right]^{-z}.
\end{equation}
The series converges for $\Re z>1/2$.
The analytical continuation of ${\cal H}(z;s,a^2)$ to the whole
complex $z$ plane is performed in 
Appendix \ref{AppendixB}.
Substituting the result into Eq.\ (\ref{49}) 
we obtain \cite{Elizalde}
\begin{eqnarray}\label{analE}
{\cal E}&=&\frac{\mu^{1-2\nu}}{8\pi}\,
\frac{\Gamma(-\nu-1)}{\Gamma(-\nu)}
\left(\frac{\pi}{\ell}\right)^{2(\nu+1)}
\left\{
\frac{1}{2}\left[(1+s)^2+\frac{g}{3}\right]^{\nu+1}
+i\int_0^{\infty}dt\,\frac{f_{\nu}(1+it)-f_{\nu}(1-it)}
{e^{2\pi t}-1}\right.
\nonumber \\
& &\left.-\frac{(1+s)^{2\nu+3}}{2\nu+3}\,F\left(-\nu-1,
-\nu-\frac{3}{2};-\nu-\frac{1}{2};-\frac{g}{3(1+s)^2}
\right)\right\},
\end{eqnarray}
where $f_{\nu}(x)\equiv\left[(x+s)^2+\frac{g}{3}\right]^{\nu+1}$
and $F(\alpha,\beta;\gamma;z)$ is the hypergeometric function.
From the definition
of the latter it follows that ${\cal E}$ has a simple pole
at $\nu=1/2$ (in fact, it has poles at
$\nu=-3/2,-1/2,1/2,3/2,\ldots$). This requires that ${\cal E}$
be renormalized before we set $\nu=1/2$. In general, this is
done by subtracting from ${\cal E}$ its value at $\ell\to\infty$.
Unfortunately, such a prescription does not work in the present case,
since, according to Eq.\ (\ref{analE}), the Casimir energy per
unit area has the form ${\cal E}=C(\nu)/\ell^{2(\nu+1)}$.

One can obtain a hint on what is going wrong by noting
that the residue of ${\cal E}$ at $\nu=1/2$ is of second order
in $g$ (or $\lambda$). This is consistent with the fact
that we have worked with the one-loop two-point Green function,
which is (formally) correct only to first
order in the coupling constant.
Since the $\lambda\Phi^4$ theory
is perturbatively renormalizable in $D=4$, one may suspect
that in order to obtain a finite (or at least renormalizable)
${\cal E}$ to order $\lambda^n$ one must work within an
approximation in which the two-point
Green function is dressed with the $n$-loop self-energy.
As we show below, this is not sufficient or necessary.
In spite of that, the argument suggests that
Eq.\ (\ref{analE}) cannot be trusted beyond order $\lambda$.

Expanding the RHS of Eq.\ (\ref{analE}) in a power
series in $\lambda$ and making $\nu\to 1/2$, we obtain
\begin{equation}\label{Eseries}
{\cal E}=\frac{1}{\ell^3}\left[-\frac{\pi^2}{1440}
+\frac{\lambda}{9216}+\ldots\right].
\end{equation}
The first term is the usual free Casimir energy (per unit area).
The second term is the leading radiative correction to it.
It overestimates the correct result \cite{Sy} by a factor
of 2. This discrepancy occurs because the method we have
used to compute the Casimir energy only works in the
absence of interactions. To show this, we start by noting
that one can define the Casimir energy as
\begin{equation}
\label{ECas}
E=\int d^{D-1}x\,\langle 0|T_{00}(x)|0\rangle,
\end{equation}
where $T_{\mu\nu}$ is the energy-momentum tensor.
In the case we are considering (the massless
$\lambda\Phi^4$ theory in $D=4$), we have
\begin{equation}
\label{T00}
T_{00}=\frac{1}{2}\left(\partial_0\Phi\right)^2
+\frac{1}{2}\left(\vec{\nabla}\Phi\right)^2
+\frac{\lambda}{4!}\,\Phi^4.
\end{equation}
Moving the differential operators outside the brackets,
we can rewrite the vacuum expectation value of $T_{00}$
in terms of $n$-point Green functions $G^{(n)}$:
\begin{equation}\label{T00(x)}
\langle 0|T_{00}(x)|0\rangle=\lim_{x'\to x}\,
\frac{1}{2}\left(\partial_0^{}\partial_0'
+\vec{\nabla}\!\cdot\!\vec{\nabla}'\right)
G^{(2)}(x,x')+\frac{\lambda}{4!}\,G^{(4)}(x,\ldots,x).
\end{equation}
On the other hand, using Eq.\ (\ref{42}) and the
spectral representation of $\widetilde{G}^{(2)}$,
Eq.\ (\ref{tG2}), one can easily show that
\begin{equation}
\frac{1}{2}\sum_{\alpha}\omega_{\alpha}=
\int d^{D-1}x\,\lim_{x'\to x}\,\frac{1}{2}
\left[\partial_0^{}\partial_0'
+\vec{\nabla}\!\cdot\!\vec{\nabla}'+\Sigma^{(1)}(x)\right]
\widetilde{G}^{(2)}(x,x').
\end{equation}
It follows that the sum of (one-loop) zero-point energies differs
from the true vacuum energy by
\begin{eqnarray}\label{DeltaE}
\Delta E&=&\int d^{D-1}x\left[\lim_{x'\to x}\frac{1}{2}
\left(\partial_0^{}\partial_0'
+\vec{\nabla}\!\cdot\!\vec{\nabla}'\right)
\Delta G^{(2)}(x,x')\right.
\nonumber \\
& &\left.-\frac{1}{2}\,\Sigma^{(1)}(x)\,
\widetilde{G}^{(2)}(x,x)
+\frac{\lambda}{4!}\,G^{(4)}(x,\ldots,x)\right],
\end{eqnarray}
where $\Delta G^{(2)}\equiv G^{(2)}-\widetilde{G}^{(2)}$.
While the first line of Eq.\ (\ref{DeltaE}) is formally
$O(\lambda^2)$, the second one is $O(\lambda)$.
This explains why the second term in Eq.\ (\ref{Eseries})
is incorrect. 

It is important to note that 
Eqs.\ (\ref{Eformal}) and (\ref{T00}) would
lead to distinct results even if we had worked with the 
complete two-point Green function. The difference
between them would then be given by
\begin{equation}
\Delta E=\int d^{D-1}x\left\{-\frac{1}{2}\int d^Dy\,
\Sigma(x,y)\,G^{(2)}(y,x)+\frac{\lambda}{4!}\,G^{(4)}(x,\ldots,x)
\right\}.
\end{equation}
A perturbative evaluation of the above expression
shows that $\Delta E$ would still be of order $\lambda$.
Physically, this discrepancy is due to the fact
that, in contrast with the free theory, the interacting
theory is not equivalent to a collection of independent
harmonic oscillators. The sum of zero-point energies,
Eq.\ (\ref{Eformal}), takes into account only the
Lamb shift on the single-particle energy levels caused
by the interaction;
the difference $\Delta E$ accounts for the residual
interaction among the (anharmonic) oscillators.

The above discussion also shows that the
Casimir energy is not determined solely by the two-point
Green function, but also (in the $\lambda\Phi^4$
theory) by the four-point function. In particular, in
order to consistently remove the $O(\lambda^2)$ UV singularity
in Eq.\ (\ref{analE}) one must not only obtain
$G^{(2)}$ to that order, but also $G^{(4)}$ to $O(\lambda)$.
These ideas will be illustrated in the next section
in the simpler case of periodic boundary conditions.


\section{Periodic boundary conditions}
\label{PBC}

\subsection{Conventional perturbation theory}
\label{conventional}

The free Feynman propagator for the field $\Phi$
obeying periodic boundary conditions in the $z$-direction,
$\Phi(t,{\bf x},z+R)=\Phi(t,{\bf x},z)$, 
is given by 
\begin{eqnarray}\label{DeltaFp}
\Delta_F(x,x')&=&\frac{i}{R}\int\frac{d\omega}{2\pi}
\sum_{n=-\infty}^{\infty}
\int\frac{d^dp}{(2\pi)^{d}}\,
\frac{e^{-ip^{\mu}(x^{}_{\mu}-x'_{\mu})}}
{\omega^2-{\bf p}^2-q_n^2-m^2+i\epsilon}
\nonumber \\
&=&\frac{1}{2R}\sum_{n=-\infty}^{\infty}
\int\frac{d^dp}{(2\pi)^{d}}\,
\frac{e^{-i\omega_n({\bf p})|t-t'|+i{\bf p}\cdot({\bf x}
-{\bf x}')+iq_n(z-z')}}{\omega_n({\bf p})}\,,
\end{eqnarray}
where $p^{\mu}=(\omega,{\bf p},q_n)$, $q_n=2\pi n/R$, and
$\omega_n({\bf p})\equiv\sqrt{{\bf p}^2+q_n^2+m^2}$.
Since $\Delta_F(x,x')=\Delta_F(x-x')$, such boundary conditions
do not break translational invariance.

The renormalized vacuum energy density may be computed from 
Eqs.\ (\ref{ECas})--(\ref{T00(x)}), but its perturbative
expansion is more easily derived from the vacuum persistence
amplitude:
\begin{equation}\label{calE}
\varepsilon=\lim_{T\to\infty}\,\frac{i}{VT}\,\ln\left[
\int{\cal D}\Phi\,\exp\left(i\int d^Dx\,{\cal L}\right)
\right]+\Lambda.
\end{equation}
The last term in Eq.\ (\ref{calE}) is fixed by the renormalization
condition $\lim_{R\to\infty}\varepsilon(x)=0$.
Due to the translational
invariance the vacuum energy density does not depend on $x$.
(A remark on notation: $\varepsilon$ denotes the 
Casimir energy per unit {\em volume}, while ${\cal E}$
denotes the Casimir energy per unit {\em area}. They
are related, in the case of periodic BC, by 
$\varepsilon={\cal E}/R$.)

To first order in $\lambda$, we obtain from Eq.\ (\ref{calE})
the well-known results [$\varepsilon=\varepsilon^{(0)}
+\varepsilon^{(1)}+\ldots$,
$\varepsilon^{(n)}=O(\lambda^n)$]
\begin{equation}\label{calE0}
\varepsilon^{(0)}
=\frac{1}{2R}\,\sum_{n=-\infty}^\infty\,\int\,
\frac{d^dp}{(2\pi)^d}\,\omega_n({\bf p})
+\Lambda^{(0)},
\end{equation}
\begin{equation}\label{calE1}
\varepsilon^{(1)}
=\frac{\lambda}{8}\left[\Delta_F (0)\right]^2+\frac{1}{2}\,
\delta m^2\,\Delta_F(0)+\Lambda^{(1)},
\end{equation}
where $\delta m^2$ is the one-loop mass counterterm.
The one-loop self-energy is given by
\begin{equation}
\Sigma^{(1)}=\frac{\lambda}{2}\,\Delta_F (0)+\delta m^2.
\end{equation}

In order to compute the quantities above, it is convenient
to define the function
\begin{equation}\label{9}
\Psi_\epsilon (\alpha)\equiv \frac{\mu^{\epsilon}}{2R}
\sum_{n=-\infty}^\infty\int
\frac{d^{d-\epsilon}p}{(2\pi)^{d-\epsilon}}\,
\left({\bf p}^2+q_n^2+m^2\right)^{\alpha},
\end{equation}
where $\mu$ is an arbitrary mass scale. We then have $\Delta_F(0)=
\lim_{\epsilon\to 0}\,\Psi_{\epsilon}(-1/2)$
and $\varepsilon^{(0)}=\lim_{\epsilon\to 0}\,
\left[\Psi_{\epsilon}(1/2) + \Lambda^{(0)}\right]$.

The computation of $\Psi_{\epsilon}$ is discussed
in Appendix \ref{AppendixC}.
There we show that $\Psi_{\epsilon}$ may be written
(in the limit $\epsilon\to 0$) as the sum of two terms, namely
\begin{equation}\label{10}
\lim_{\epsilon\to 0}\,\Psi_\epsilon(\alpha)={\cal A}(\alpha)+
{\cal B}(\alpha),
\end{equation}
where ${\cal A}(\alpha)$ and ${\cal B}(\alpha)$
are given by Eqs.\ (\ref{calA}) and (\ref{calB}), respectively.
Only ${\cal A}(\alpha)$ depends on $R$, and vanishes when
$R\to \infty$.

Before computing $\varepsilon^{(0)}$ and $\varepsilon^{(1)}$
we give our renormalization conditions. To first order
in $\lambda$
two conditions are required. We fix $\Lambda$ and
$\delta m^2$ by the conditions
$\lim_{R\to\infty}\,\varepsilon(R)=0$ and
$\lim_{R\to\infty}\,\Sigma^{(1)}(R)=0$, respectively.
This gives
$\Lambda^{(0)}=-{\cal B}(1/2)$,
$\Lambda^{(1)}=\frac{\lambda}{8}
\left[ {\cal B}(-1/2)\right]^2$, and
$\delta m^2=-\frac{\lambda}{2}\,{\cal B}(-1/2)$.
It follows that ($\epsilon\to 0$)
\begin{equation}\label{11}
\varepsilon^{(0)}(R)={\cal A}(1/2)=
-\frac{2}{(2\pi)^{D/2}}
\left(\frac{m}{R}\right)^{D/2}
F\left(\frac{D}{2};mR\right),
\end{equation}
\begin{equation}\label{13}
\varepsilon^{(1)}(R)=\frac{\lambda}{8}
\left[{\cal A}(-1/2)\right]^2
=\frac{\lambda}{2(2\pi)^{D}}
\left(\frac{m}{R}\right)^{D-2}\left[
F\left(\frac{D}{2}-1;mR\right)\right]^2,
\end{equation}
where $F(s;a)$ is defined in Eq.\ (\ref{F(s;a)}).
Taking $D=4$ and expanding in powers of $mR$ [Eqs.\ (\ref{A7})
and (\ref{A8})] we thus obtain
\begin{equation}\label{14}
\varepsilon=\frac{1}{R^4}\left\{\left[-\frac{\pi^2}{90}
+\frac{(mR)^2}{24}-\frac{(mR)^3}{12\pi}+\ldots\right]
+\lambda\left[\frac{1}{1152}
-\frac{mR}{192\pi}+\ldots\right]\right\}+\ldots
\end{equation}
Analogously, we obtain for the one-loop self-energy
\begin{eqnarray}\label{12b}
\Sigma^{(1)}&=&\frac{\lambda}{(2\pi)^{D/2}}
\left(\frac{m}{R}\right)^{(D-2)/2}
F\left(\frac{D}{2}-1;mR\right)
\nonumber\\
&=&\frac{\lambda}{R^2}\left[\frac{1}{24}
-\frac{mR}{8\pi}+\frac{(mR)^2}{16\pi^2}\,
\ln\left(mR\right)+\ldots\right]\qquad(D=4).
\end{eqnarray}
The first term
does not depend on $m$, and is sometimes called
``topological mass'' (squared). For reasons
discussed in \cite{Fulling}, we prefer the name
``compactification mass'' for $M\equiv(\lambda/24R^2)^{1/2}$.


\subsection{Resummed perturbation theory}
\label{resummed}

From now on, let us focus the discussion on the massless
theory ($m=0$) in $D=4$. In this case, the second and
higher order terms in the perturbative
expansion of $\varepsilon$ are plagued with IR divergences.
A qualitative analysis shows that the most IR divergent diagrams are
the ``ring'' (or ``daisy'') ones.
(These are just the diagrams with the greatest number of
insertions of the one-loop self-energy in each
order of perturbation theory.)
As in the case of TFT \cite{Thermal}, it is possible to sum these
diagrams to all orders. The result is finite
in the IR and is nonanalytic in $\lambda$, as we show
below.

To avoid overcounting of diagrams in higher order
calculations, it is convenient to redefine the free
and the interacting parts of the Lagrangian by adding and
subtracting to it the compactification mass term
$\frac{1}{2}\,M^2\Phi^2$ \cite{Braaten,RRParwani,Frenkel}:
\begin{eqnarray}\label{20}
{\cal L}&=&\widetilde{\cal L}_{0} + \widetilde{\cal L}_{I}
\nonumber \\
&=&\left\{\frac{1}{2}\,(\partial\Phi)^2-\frac{1}{2}\,M^2
\Phi^2\right\}
+\left\{-\frac{\lambda}{4!}\,\Phi^4+ \frac{1}{2}\,
M^2\Phi^2+{\cal L}_{\rm ct}\right\}.
\end{eqnarray}
The free propagator (in momentum space) is now given by
\begin{equation}
\widetilde{\Delta}_F(p)=\frac{i}{p^2-M^2+i\epsilon}.
\end{equation}
It coincides with the propagator of the original
theory in the daisy approximation.

We remark that in a loop expansion of the vacuum energy
(or of any other quantity) each insertion of the mass
term in $\widetilde{\cal L}_I$ is to be formally counted
as one loop, like the mass counterterm --- otherwise
taking $\widetilde{\cal L}_0$ as the new
free Lagrangian would not cure the IR divergence
problem \cite{Coleman}.

The one-loop approximation to the vacuum energy is
given by
\begin{equation}\label{23b}
\tilde\varepsilon^{(1)}
=\frac{1}{2R}\sum_{n=-\infty}^{\infty}\int
\frac{d^2p}{(2\pi)^2}\left({\bf p}^{2}+q_n^2+M^2\right)^{1/2}.
\end{equation}
Using the results of Section \ref{conventional} and of
Appendix \ref{AppendixC} we obtain
\begin{eqnarray}\label{24}
\tilde\varepsilon^{(1)}(R)&=&-\frac{M^2}{2\pi^2R^2}\,
F(2;MR)+\lim_{\epsilon\to 0}\,
\frac{\mu^{\epsilon}M^{4-\epsilon}}{2^{4-\epsilon}
\pi^{(3-\epsilon)/2}}\,\frac{\Gamma\left(-2+\frac{\epsilon}{2}
\right)}{\Gamma\left(-\frac{1}{2}\right)}
\nonumber \\
&=&\frac{1}{R^4}\left[-\frac{\pi^2}{90}
+\frac{\lambda}{576}
-\frac{\lambda^{3/2}}{288\sqrt{6}\pi}
+O\left(\frac{\lambda^2}{\epsilon}\right)\right].
\end{eqnarray}

As in the Dirichlet BC case, the $O(\lambda)$ term in
the one-loop approximation is twice the value
obtained in conventional perturbation theory
[Eq.\ (\ref{14}) with $m=0$]. In order to reproduce the
latter one has to take into account the two-loop contribution
to $\varepsilon$, given by
\begin{equation}\label{25}
\tilde\varepsilon^{(2)}
=\frac{\lambda}{8}\left[\widetilde{\Delta}_{F}(0)\right]^2
-\frac{1}{2}\,M^2\,\widetilde{\Delta}_{F}(0),
\end{equation}
where $\widetilde{\Delta}_F(x)=\Delta_F(x;m=M)$.
Using again results of Section \ref{conventional} and of
Appendix \ref{AppendixC} we obtain
\begin{eqnarray}
\widetilde{\Delta}_F(0)&=&\frac{M}{2\pi^2R}\,F(1;MR)
+\lim_{\epsilon\to 0}\,\frac{\mu^{\epsilon}
M^{2-\epsilon}}{2^{4-\epsilon}\pi^{(3-\epsilon)/2}}
\,\frac{\Gamma\left(-1+\frac{\epsilon}{2}\right)}
{\Gamma\left(\frac{1}{2}\right)}
\nonumber \\
&=&\frac{1}{R^2}\left[\frac{1}{12}
-\frac{\lambda^{1/2}}{8\pi\sqrt{6}}
+O\left(\frac{\lambda}{\epsilon}\right)\right].
\end{eqnarray}
Substituting this into Eq.\ (\ref{25}) yields
\begin{equation}\label{26}
\tilde\varepsilon^{(2)}(R)=\frac{1}{R^4}
\left[-\frac{\lambda}{1152}+O\left(\frac{\lambda^2}
{\epsilon}\right)\right].
\end{equation}
Thus, to order $\lambda^2$ we finally obtain
\begin{equation}\label{Eres}
\varepsilon(R)=\frac{1}{R^4}\left[-\frac{\pi^2}{90}
+\frac{\lambda}{1152}
-\frac{\lambda^{3/2}}{288\sqrt{6}\pi}
+O\left(\frac{\lambda^2}{\epsilon}\right)\right].
\end{equation}
This agrees to order $\lambda$ with the result found in
Section \ref{conventional} (in the limit $m\to 0$).
Besides, we have obtained a correction of order $\lambda^{3/2}$.
This nonanalyticity in $\lambda$ is a consequence of the fact
that the loop expansion in the rearranged Lagrangian is
equivalent to a resummation of an infinite number of graphs
in the conventional perturbation expansion.

Finally, we note that the UV singularities in the
resummed theory depend on $R$, via their dependence on $M$.
For instance, in the computation of $\tilde\varepsilon^{(1)}$
a singular term of the form 
$M^4/\epsilon\sim\lambda^2/\epsilon R^4$ appears in the
limit $\epsilon\to 0$. In the analogous
case of TFT it can be shown that the UV singularity
present in the one-loop free energy
cancels against two- and three-loop
contributions in the resummed theory, including
a coupling constant renormalization 
counterterm \cite{Frenkel,Parwani}.
These contributions on their turn introduce
new singularities at $O(\lambda^3)$, which are cancelled
by including higher order graphs in the resummed theory,
and so on. 
The situation is exactly the same in our case, so
we can safely neglect the $O(\lambda^2)$ term in
Eq.\ (\ref{Eres}).


\section{Conclusion}

In this paper we have discussed the computation of 
radiative corrections to the Casimir energy of the
massless $\lambda\Phi^4$ theory confined between two
parallel plates. The case of Dirichlet boundary conditions
at the plates was discussed in Section \ref{DBC}.
We obtained an analytical expression for the one-loop 
self-energy $\Sigma^{(1)}(x)$ both in $D=3$ and $D=4$.
The former was shown to be free of IR singularities,
in contrast with the claim made in \cite{Nami}. 
 
In the ``daisy'' resummation of the two-point Green function
one is led to solve a Klein-Gordon equation 
with $\Sigma^{(1)}(x)$ acting as an effective 
scalar potential. We were able to solve this equation
in four dimensions. In spite of $\Sigma^{(1)}$ being
negative everywhere, there are
no tachyonic modes if the coupling constant $\lambda$
is smaller than $\lambda_{\rm crit}=8\pi^2$.
We then computed the sum of the eigenenergies of 
the Klein-Gordon operator. Expanding the result in a power
series in $\lambda$ one discovers that the 
$O(\lambda)$ correction does not agree with the 
result of conventional perturbation theory,
and the correction of order $\lambda^2$
contains a UV singularity which apparently 
cannot be renormalized away. The first problem was
shown to occur because the sum of zero-point energies
does not take into account all the contributions
to the vacuum energy in a theory with interaction.
As for the second problem, it was argued that the consistent
renormalization of the Casimir energy at a given order
requires that one takes into account all diagrams to that order.
This conjecture is supported by the fact that
the Dirichlet BC (in the case of flat boundaries)
do not spoil the perturbative renormalizability of the 
$\lambda\Phi^4_4$ theory \cite{Sy}. 

In the case of periodic boundary conditions 
in one spatial direction we have argued that the infrared
properties of the model
are analogous to the one in thermal field theory. In order to define
a consistent (i.e., IR finite) perturbative expansion 
one has to include the screening effects due to collective
excitations. A solution to this
problem was proposed by Braaten, Pisarski and others 
in thermal field theory
\cite{Braaten,RRParwani,Frenkel}. It consists in the resummation of
an infinite class of diagrams, which gives the field an 
effective mass. This can be done in a systematic way using
the Braaten-Pisarski resummation method. 
This was illustrated with the calculation of the leading and
next-to-leading order radiative corrections to the 
Casimir energy. Besides, our calculation shows that
the resummed weak coupling expansion of the Casimir energy 
in the case of periodic BC
contains fractional powers of $\lambda$, in contrast to the
case of Dirichlet boundary conditions. 

We note that calculations of radiative 
corrections to the Casimir
energy via the resummation of zero-point energies have appeared
recently in the literature \cite{Nam},
without paying due attention to the subtleties
of the resummed perturbation theory. 
As we have shown, this may lead to inconsistencies 
in the results \cite{Albuquerque}.

Finally, we hope that the techniques discussed here may be
useful in investigations of Kaluza-Klein compactification
scenarios, as well as in the study of surface critical phenomena.


\acknowledgments

The authors acknowledge the financial support from FAPESP under
grants 00/03277-3 (L.C.A.) and 98/11646-7 (R.M.C.), 
and from FAPERJ (R.M.C.).
They also acknowledge the kind hospitality of the Departamento de
F\'\i sica Matem\'atica, Universidade de S\~ao Paulo,
where this work was initiated.


\appendix

\section{}
\label{AppendixA}

We discuss the solution to the equation
\begin{equation}\label{B1}
\left(\frac{d^2}{dy^2}+k^2+\frac{g}{\sin^2y}\right)
\varphi(y)=0,
\end{equation}
with Dirichlet boundary conditions at $y=0$ and $y=\pi$.

Let us first
consider the asymptotic behavior of its solutions near one of
the boundaries (say, at $y=0$). To this end, we can replace
Eq.\ (\ref{B1}) by
\begin{equation}\label{B10}
\left(\frac{d^2}{dy^2}+\frac{g}{y^2}\right)
\varphi(y)=0.
\end{equation}
The most general solution to Eq.\ (\ref{B10}) is
\begin{equation}\label{phi_as}
\varphi(y)=A\,y^{s_+}+B\,y^{s_-},
\end{equation}
where
$s_{\pm}\equiv\frac{1}{2}\left(1\pm\sqrt{1-4g}\right)$.
If $g<1/4$, the boundary condition $\varphi(0)=0$ is not
enough to fix the relation between $A$ and $B$, as both
$y^{s_+}$ and $y^{s_-}$ vanish at $y=0$. To resolve this
indeterminacy, we follow \cite{Landau} and regularize the
potential near the origin: $V_R(y)=-g/y^2$ for $y>a$,
and $V_R(y)=-g/a^2$ for $y<a$. At the end, we shall
take the limit $a\to 0$.

For $y>a$, the solution is given by Eq.\ (\ref{phi_as}).
For $y<a$, the solution which satisfies the boundary condition
at the origin is $\varphi(y)=C\,\sin\left(\sqrt{g}y/a\right)$.
Continuity of $\varphi(y)$
and its derivative at $y=a$ implies the relation
$B/A\sim a^{s_+ - s_-}$ as $a\to 0$, i.e., only
the solution with the faster decay at the origin survives
when the regularization is removed.
If $g>1/4$, $s_+ - s_-$ is purely imaginary and
$\lim_{a\to 0}B/A$ does not exist. This sets a critical value to
$g$, namely $g_{\rm crit}=1/4$,
above which the ``Hamiltonian'' $H=-d^2/dy^2-g/y^2$
is unbounded from below \cite{Landau}.

Let us return to the complete equation (\ref{B1}).
It is convenient to make some changes of variables.
First, we set $y=ix+\pi/2$ and define
$s\equiv\frac{1}{2}\left(-1+\sqrt{1-4g}\right)$
[so that $s(s+1)=-g$]. This transforms Eq.\ (\ref{B1})
into
\begin{equation}
\left[-\frac{d^2}{dx^2}+k^2-\frac{s(s+1)}{\cosh^2 x}\right]
\varphi(x)=0.
\end{equation}
Then, we make $\xi=\tanh x$ and obtain
\begin{equation}\label{B2}
\frac{d}{d\xi}\left[(1-\xi^2)\,\frac{d\varphi}{d\xi}\right]
+\left[s(s+1)-\frac{k^2}{1-\xi^2}\right]\varphi(\xi)=0.
\end{equation}
Finally, we put $\varphi(\xi)=(1-\xi^2)^{k/2}\,w(\xi)$, followed by
the change of variable $\xi=2u-1$, to get
\begin{equation}\label{B3}
u(1-u)\,\frac{d^2w}{du^2}+[1+k-2(1+k)u]\,\frac{dw}{du}
-(k-s)(k+s+1)\,w=0.
\end{equation}

Eq.\ (\ref{B3}) is the hypergeometric differential
equation \cite{Grad} with
parameters $\alpha=k-s$, $\beta=k+s+1$, and $\gamma=1+k$.
Its general solution may be written as
\begin{equation}\label{B4}
w(u)=A\,(1-u)^{-k}\,F(-s,s+1;1+k;u)
+B\,u^{-k}\,F(-s,s+1;1-k;u),
\end{equation}
where $F(\alpha,\beta;\gamma;z)$ is the hypergeometric
function. Returning to the variable $y$ and the function
$\varphi$, we have
\begin{equation}\label{B6}
\varphi(y)=A'\,e^{-iky}
F\left(-s,s+1;1+k;\frac{i\,e^{-iy}}{2\,\sin y}\right)
+B'\,e^{iky}
F\left(-s,s+1;1-k;\frac{i\,e^{-iy}}{2\,\sin y}\right).
\end{equation}
The asymptotic behavior of $\varphi(y)$ as $y\to 0$
may be extracted from
$\lim_{z\to 0}\,F(\alpha,\beta;\gamma;z)=1$, after using
the relation
(valid for $|{\rm arg}(-z)|<\pi$, $|{\rm arg}(1-z)|<\pi$,
$\alpha-\beta\ne 0,\pm 1,\pm 2,\ldots$)
\begin{eqnarray}\label{B7}
F(\alpha,\beta;\gamma;z)&=&(-z)^{-\alpha}\,
\frac{\Gamma(\gamma)\,\Gamma(\beta-\alpha)}
{\Gamma(\gamma-\alpha)\,\Gamma(\beta)}\,
F\left(\alpha,1+\alpha-\gamma;1+\alpha-\beta;\frac{1}{z}\right)
\nonumber \\
& &+(-z)^{-\beta}\,
\frac{\Gamma(\gamma)\,\Gamma(\alpha-\beta)}
{\Gamma(\gamma-\beta)\,\Gamma(\alpha)}\,
F\left(\beta,1+\beta-\gamma;1+\beta-\alpha;\frac{1}{z}\right).
\end{eqnarray}
In this way,
\begin{eqnarray}\label{B8}
\varphi(y)&\stackrel{y\to 0}{\sim}&
A'\,F\left(-s,s+1;1+k;\frac{i}{2y}\right)
+B'\,F\left(-s,s+1;1-k;\frac{i}{2y}\right)
\nonumber \\
&\sim&\left[A'\,\frac{\Gamma(1+k)}{\Gamma(1+k+s)}
+B'\,\frac{\Gamma(1-k)}{\Gamma(1-k+s)}\right]
\frac{\Gamma(2s+1)}{\Gamma(s+1)}
\left(-\frac{i}{2y}\right)^s
\nonumber \\
& &+\left[A'\,\frac{\Gamma(1+k)}{\Gamma(k-s)}
+B'\,\frac{\Gamma(1-k)}{\Gamma(-k-s)}\right]
\frac{\Gamma(-2s-1)}{\Gamma(-s)}
\left(-\frac{i}{2y}\right)^{-s-1}.
\end{eqnarray}
Recalling the analysis of Eq.\ (\ref{B10}),
we impose that $\varphi(y)\sim y^{s+1}$ as $y\to 0$.
This can be accomplished by taking
$B'=0$ and $k=-(n+s)$ ($n=1,2,\ldots$), that is
\begin{equation}\label{B9}
\varphi(y)=A'\,e^{i(n+s)y}F\left(-s,s+1;-s-n+1;
\frac{i\,e^{-iy}}{2\,\sin y}\right).
\end{equation}
The boundary condition at $y=\pi$ is also
satisfied by this solution, since
$\varphi(y)\sim(\pi-y)^{s+1}$ as $y\to\pi$. Hence,
Eq.\ (\ref{B9}) is an acceptable solution to Eq.\ (\ref{B1}).
[Remark: we could also have taken $A'=0$ and $k=n+s$
($n=1,2,\ldots$) in Eq.\ (\ref{B8}), but this leads
to the same values of $k^2$ and to the same solutions
given by Eq.\ (\ref{B9}).]


\section{}
\label{AppendixB}

The goal here is to obtain the analytical continuation of
the function ${\cal H}(z;s,a^2)$, defined in Eq.\ (\ref{calH}),
to the whole complex $z-$plane. 
To this end, we shall use the Plana summation formula \cite{Whit}
\begin{eqnarray}\label{C2}
\sum_{k=M}^N\,f(k)&=&
\frac{1}{2}\left[f(M)+f(N)\right]+\int_M^N\,f(x)dx
\nonumber\\
& &-i\int_0^{\infty} dy\,\frac{f(N+iy)-f(M+iy)-f(N-iy)+f(M-iy)}
{e^{2\pi y}-1}.
\end{eqnarray}

In the present case, we choose $M=1$, $N=\infty$, and
$f(x)=\left[(x+s)^2+a^2\right]^{-z}$. In order to apply
the Plana formula some conditions have to be satisfied \cite{Whit}.
First, we assume that $\Re z>\frac{1}{2}$, so that
the series in Eq.\ (\ref{calH}) converges absolutely.
Then, it can be shown that the function $f(\tau+it)$ is
holomorphic for $\tau\geq1$ for any $t$, and that
$\lim_{t\to\pm\infty}\,e^{-2\pi|t|}\,f(\tau+it)=0$
uniformly in the interval $1\leq\tau<\infty$.
In addition, $\lim_{\tau\to\infty}\int_{-\infty}^\infty
dt\,e^{-2\pi|t|}|f(\tau+it)|=0$. Under these conditions, we have
\begin{eqnarray}\label{C3}
{\cal H}(z;s,a^2)&=&\frac{1}{2}\left[(s+1)^2+a^2\right]^{-z}
+\int_1^\infty\frac{dx}{\left[(x+s)^2+a^2\right]^{z}}
\nonumber \\
& &+i\int_0^\infty dt\,\frac{f(1+it)-f(1-it)}{e^{2\pi t}-1}.
\end{eqnarray}
The first integral can be computed in closed form:
\begin{equation}\label{C5}
\int_1^\infty\frac{dx}{\left[(x+s)^2+a^2\right]^{z}}
=\frac{(1+s)^{1-2z}}{2z-1}\,F\left(z,z-\frac{1}{2};
z+\frac{1}{2};-\left(\frac{a}{1+s}\right)^2\right).
\end{equation}
Eqs.\ (\ref{C3}) and (\ref{C5}) give the analytic continuation
of ${\cal H}(z;s,a^2)$ to the whole complex $z$-plane. From
the definition of the hypergeometric function it follows
that ${\cal H}(z;s,a^2)$ has simple poles at
$z=1/2,-1/2,-3/2,\ldots$


\section{}
\label{AppendixC}

Consider the function
\begin{equation}\label{A1}
\Psi_\epsilon (\alpha)\equiv \frac{\mu^{\epsilon}}{2R}
\sum_{n=-\infty}^{\infty}\int\frac{d^{d-\epsilon}p}
{(2\pi)^{d-\epsilon}}\,
\left[p^2+\left(\frac{2\pi n}{R}\right)^2+m^2\right]^{\alpha}.
\end{equation}
Integration over $p$ leads to
\begin{equation}\label{A2}
\Psi_\epsilon(\alpha)=\frac{\mu^{\epsilon}}{2^{d+1-\epsilon}R}\,
\frac{\pi^{\alpha}}{\Gamma(-\alpha)}\,
S\left(m,\frac{R}{2};-2\alpha-d+\epsilon\right),
\end{equation}
where
\begin{equation}\label{A3}
S(m,a;s)\equiv\pi^{-s/2}\,\Gamma\left(\frac{s}{2}\right)
\sum_{n=-\infty}^{\infty}\left[\left(\frac{m}{\pi}\right)^2
+\left(\frac{n}{a}\right)^2\right]^{-s/2}.
\end{equation}
The series converges absolutely for $\Re s> 1$.
The analytical continuation to a meromorphic function in the complex $s$
plane is given by \cite{Ambjorn}
\begin{equation}\label{A4}
S(m,a;s)=\frac{am^{1-s}}{\pi^{(1-s)/2}}\left[
\Gamma\left(\frac{s-1}{2}\right)+4(ma)^{(s-1)/2}\,
F\left(\frac{1-s}{2};2ma\right)\right],
\end{equation}
where
\begin{equation}\label{F(s;a)}
\quad F(s;a)\equiv\sum_{n=1}^{\infty}n^{-s}K_{s}(na),
\end{equation}
with $K_{s}(x)$ the modified Bessel function of the second kind.
The following expansions, valid for $a\ll 1$, will be 
useful \cite{Braden}:
\begin{equation}\label{A7}
F(1;a)=\frac{\pi^2}{6a}-\frac{\pi}{2}+O(a\ln a),
\end{equation}
\begin{equation}\label{A8}
F(2;a)=\frac{\pi^4}{45a^2}-\frac{\pi^2}{12}+\frac{\pi a}{6}+
O(a^2).
\end{equation}

Using the analytical continuation given in Eq.\ (\ref{A4}), one may
write
$\Psi_\epsilon(\alpha)$ (in the limit $\epsilon\to 0$)
as the sum of two terms, namely,
$\lim_{\epsilon\to 0}\,\Psi_{\epsilon}(\alpha)={\cal A}(\alpha)
+{\cal B}(\alpha)$, where
\begin{equation}\label{calA}
{\cal A}(\alpha)=\frac{2^{(1+2\alpha-d)/2}}{\pi^{(1+d)/2}\,
\Gamma(-\alpha)}\left(\frac{m}{R}\right)^{(1+2\alpha+d)/2}
F\left(\frac{1+2\alpha+d}{2};mR\right),
\end{equation}
and
\begin{equation}\label{calB}
{\cal B}(\alpha)=\lim_{\epsilon\to 0}\,\frac{\mu^{\epsilon}\,
m^{1+2\alpha+d-\epsilon}}{2^{d+2-\epsilon}\,\pi^{(1+d-\epsilon)/2}}\,
\frac{\Gamma\left(\frac{-1-2\alpha-d+\epsilon}{2}\right)}
{\Gamma(-\alpha)}.
\end{equation}
Note that only ${\cal A}(\alpha)$ depends on $R$.


\end{document}